\pdfoutput=1
\documentclass[12pt,a4paper]{article}
\usepackage{siunitx}

\usepackage{ifthen} 
\newboolean{pdflatex}
\setboolean{pdflatex}{true} 

\newboolean{articletitles}
\setboolean{articletitles}{true}

\newboolean{uprightparticles}
\setboolean{uprightparticles}{false} 

\newboolean{inbibliography}
\setboolean{inbibliography}{false} 

\def\paperauthors{LHCb collaboration} 
\def\paperasciititle{Observation of the decay  Lb -> Lc p pbar pi} 
\def\papertitle{Observation of the decay $\Lb \to \Lc  p \overline{p} \pi^-$} 
\def\paperkeywords{{High Energy Physics}, {LHCb}} 
\def\papercopyright{\the\year\ CERN for the benefit of the LHCb collaboration} 
\def\paperlicence{CC-BY-4.0 licence}
\def\paperlicenceurl{https://creativecommons.org/licenses/by/4.0/}

\usepackage[top=1in, bottom=1.25in, left=1in, right=1in]{geometry}

\usepackage{microtype}
\usepackage{lineno}  
\usepackage{xspace} 
\usepackage{caption}

\usepackage{graphicx}  
\usepackage{color}
\usepackage{colortbl}
\graphicspath{{./}} 

\usepackage{amsmath} 
\usepackage{amssymb}
\usepackage{amsfonts}
\usepackage{upgreek} 

\newcommand*\patchAmsMathEnvironmentForLineno[1]{%
\expandafter\let\csname old#1\expandafter\endcsname\csname #1\endcsname
\expandafter\let\csname oldend#1\expandafter\endcsname\csname
end#1\endcsname
 \renewenvironment{#1}%
   {\linenomath\csname old#1\endcsname}%
   {\csname oldend#1\endcsname\endlinenomath}%
}
\newcommand*\patchBothAmsMathEnvironmentsForLineno[1]{%
  \patchAmsMathEnvironmentForLineno{#1}%
  \patchAmsMathEnvironmentForLineno{#1*}%
}
\AtBeginDocument{%
\patchBothAmsMathEnvironmentsForLineno{equation}%
\patchBothAmsMathEnvironmentsForLineno{align}%
\patchBothAmsMathEnvironmentsForLineno{flalign}%
\patchBothAmsMathEnvironmentsForLineno{alignat}%
\patchBothAmsMathEnvironmentsForLineno{gather}%
\patchBothAmsMathEnvironmentsForLineno{multline}%
\patchBothAmsMathEnvironmentsForLineno{eqnarray}%
}

\usepackage{hyperxmp}

\usepackage[pdftex,
            pdfauthor={\paperauthors},
            pdftitle={\paperasciititle},
            pdfkeywords={\paperkeywords},
            pdfcopyright={Copyright (C) \papercopyright},
            pdflicenseurl={\paperlicenceurl}]{hyperref}

\usepackage[all]{hypcap} 

\usepackage{xspace} 
\usepackage{upgreek}

\def\lhcb {\mbox{LHCb}\xspace}

\def\MagUp {\mbox{\em Mag\kern -0.05em Up}\xspace}

\ifthenelse{\boolean{uprightparticles}}%
{

 \def\Ppi         {\ensuremath{\uppi}\xspace}                 
                  
 \def\Prho        {\ensuremath{\uprho}\xspace}

 \def\Ppsi        {\ensuremath{\uppsi}\xspace}

 \def\PDelta      {\ensuremath{\Delta}\xspace}                 
 \def\PXi      {\ensuremath{\Xi}\xspace}                 
 \def\PLambda      {\ensuremath{\Lambda}\xspace}                 
 \def\PSigma      {\ensuremath{\Sigma}\xspace}                 
 \def\POmega      {\ensuremath{\Omega}\xspace}                 
 \def\PUpsilon      {\ensuremath{\Upsilon}\xspace}                 
 
 \def\PB      {\ensuremath{\mathrm{B}}\xspace}                 
                  
 \def\PD      {\ensuremath{\mathrm{D}}\xspace}

 \def\PJ      {\ensuremath{\mathrm{J}}\xspace}                 
 \def\PK      {\ensuremath{\mathrm{K}}\xspace}

 \def\Pb      {\ensuremath{\mathrm{b}}\xspace}                 
 \def\Pc      {\ensuremath{\mathrm{c}}\xspace}

 \def\Pi      {\ensuremath{\mathrm{i}}\xspace}

 \def\Pp      {\ensuremath{\mathrm{p}}\xspace}                 
 \def\Pq      {\ensuremath{\mathrm{q}}\xspace}                 
                  
 \def\Ps      {\ensuremath{\mathrm{s}}\xspace}

}
{

 \def\Ppi         {\ensuremath{\pi}\xspace}                 
                  
 \def\Prho        {\ensuremath{\rho}\xspace}

 \def\Ppsi        {\ensuremath{\psi}\xspace}                 
                  
 \mathchardef\PDelta="7101
 \mathchardef\PXi="7104
 \mathchardef\PLambda="7103
 \mathchardef\PSigma="7106
 \mathchardef\POmega="710A
 \mathchardef\PUpsilon="7107
                  
 \def\PB      {\ensuremath{B}\xspace}                 
                  
 \def\PD      {\ensuremath{D}\xspace}

 \def\PJ      {\ensuremath{J}\xspace}                 
 \def\PK      {\ensuremath{K}\xspace}

 \def\Pb      {\ensuremath{b}\xspace}                 
 \def\Pc      {\ensuremath{c}\xspace}

 \def\Pi      {\ensuremath{i}\xspace}

 \def\Pp      {\ensuremath{p}\xspace}                 
 \def\Pq      {\ensuremath{q}\xspace}                 
                  
 \def\Ps      {\ensuremath{s}\xspace}

}

\makeatletter
\ifcase \@ptsize \relax
  \newcommand{\miniscule}{\@setfontsize\miniscule{4}{5}}
\or
  \newcommand{\miniscule}{\@setfontsize\miniscule{5}{6}}
\or
  \newcommand{\miniscule}{\@setfontsize\miniscule{5}{6}}
\fi
\makeatother

\DeclareRobustCommand{\optbar}[1]{\shortstack{{\miniscule (\rule[.5ex]{1.25em}{.18mm})}
  \\ [-.7ex] $#1$}}

\def\quark     {{\ensuremath{\Pq}}\xspace}
\def\quarkbar  {{\ensuremath{\overline \quark}}\xspace}
\def\qqbar     {{\ensuremath{\quark\quarkbar}}\xspace}

\def\squark    {{\ensuremath{\Ps}}\xspace}

\def\cquark    {{\ensuremath{\Pc}}\xspace}

\def\bquark    {{\ensuremath{\Pb}}\xspace}

\def\pion   {{\ensuremath{\Ppi}}\xspace}
\def\piz    {{\ensuremath{\pion^0}}\xspace}

\def\pip    {{\ensuremath{\pion^+}}\xspace}
\def\pim    {{\ensuremath{\pion^-}}\xspace}

\def\rhomeson {{\ensuremath{\Prho}}\xspace}

\def\rhom     {{\ensuremath{\rhomeson^-}}\xspace}

\def\kaon    {{\ensuremath{\PK}}\xspace}
  \def\Kbar    {{\kern 0.2em\overline{\kern -0.2em \PK}{}}\xspace}

\def\KorKbar    {\kern 0.18em\optbar{\kern -0.18em K}{}\xspace}

\def\Kp      {{\ensuremath{\kaon^+}}\xspace}
\def\Km      {{\ensuremath{\kaon^-}}\xspace}

  \def\Dbar    {{\kern 0.2em\overline{\kern -0.2em \PD}{}}\xspace}
\def\D       {{\ensuremath{\PD}}\xspace}

\def\DorDbar    {\kern 0.18em\optbar{\kern -0.18em D}{}\xspace}

\def\Dp      {{\ensuremath{\D^+}}\xspace}

\def\Dsp     {{\ensuremath{\D^+_\squark}}\xspace}

\def\Bbar    {{\ensuremath{\kern 0.18em\overline{\kern -0.18em \PB}{}}}\xspace}

\def\BorBbar    {\kern 0.18em\optbar{\kern -0.18em B}{}\xspace}

\def\Bzb     {{\ensuremath{\Bbar{}^0}}\xspace}

\def\Bsb     {{\ensuremath{\Bbar{}^0_\squark}}\xspace}

\def\jpsi     {{\ensuremath{{\PJ\mskip -3mu/\mskip -2mu\Ppsi\mskip 2mu}}}\xspace}

  \def\Y#1S{\ensuremath{\PUpsilon{(#1S)}}\xspace}

\def\proton      {{\ensuremath{\Pp}}\xspace}
\def\antiproton  {{\ensuremath{\overline \proton}}\xspace}

\def\Lz          {{\ensuremath{\PLambda}}\xspace}
\def\Lbar        {{\ensuremath{\kern 0.1em\overline{\kern -0.1em\PLambda}}}\xspace}
\def\LorLbar    {\kern 0.18em\optbar{\kern -0.18em \PLambda}{}\xspace}

\def\Lb      {{\ensuremath{\Lz^0_\bquark}}\xspace}

\def\Lc      {{\ensuremath{\Lz^+_\cquark}}\xspace}

\def\to                 {\ensuremath{\rightarrow}\xspace}

\def\AT#1     {\ensuremath{A_{\mathrm{T}}^{#1}}\xspace}           

\def\C#1      {\ensuremath{\mathcal{C}_{#1}}\xspace}                       
\def\Cp#1     {\ensuremath{\mathcal{C}_{#1}^{'}}\xspace}                    
\def\Ceff#1   {\ensuremath{\mathcal{C}_{#1}^{\mathrm{(eff)}}}\xspace}        
\def\Cpeff#1  {\ensuremath{\mathcal{C}_{#1}^{'\mathrm{(eff)}}}\xspace}       
\def\Ope#1    {\ensuremath{\mathcal{O}_{#1}}\xspace}                       
\def\Opep#1   {\ensuremath{\mathcal{O}_{#1}^{'}}\xspace}                    

\newcommand{\tev}{\ifthenelse{\boolean{inbibliography}}{\ensuremath{~T\kern -0.05em eV}}{\ensuremath{\mathrm{\,Te\kern -0.1em V}}}\xspace}
\newcommand{\gev}{\ensuremath{\mathrm{\,Ge\kern -0.1em V}}\xspace}
\newcommand{\mev}{\ensuremath{\mathrm{\,Me\kern -0.1em V}}\xspace}
\newcommand{\kev}{\ensuremath{\mathrm{\,ke\kern -0.1em V}}\xspace}
\newcommand{\ev}{\ensuremath{\mathrm{\,e\kern -0.1em V}}\xspace}
\newcommand{\gevc}{\ensuremath{{\mathrm{\,Ge\kern -0.1em V\!/}c}}\xspace}
\newcommand{\mevc}{\ensuremath{{\mathrm{\,Me\kern -0.1em V\!/}c}}\xspace}
\newcommand{\gevcc}{\ensuremath{{\mathrm{\,Ge\kern -0.1em V\!/}c^2}}\xspace}
\newcommand{\gevgevcccc}{\ensuremath{{\mathrm{\,Ge\kern -0.1em V^2\!/}c^4}}\xspace}
\newcommand{\mevcc}{\ensuremath{{\mathrm{\,Me\kern -0.1em V\!/}c^2}}\xspace}

\def\fm   {\ensuremath{\mathrm{ \,fm}}\xspace}

\def\invfb   {\ensuremath{\mbox{\,fb}^{-1}}\xspace}

\def\ps   {\ensuremath{{\mathrm{ \,ps}}}\xspace}

\newcommand{\chisq}{\ensuremath{\chi^2}\xspace}

\newcommand{\chisqip}{\ensuremath{\chi^2_{\text{IP}}}\xspace}

\def\gsim{{~\raise.15em\hbox{$>$}\kern-.85em
          \lower.35em\hbox{$\sim$}~}\xspace}
\def\lsim{{~\raise.15em\hbox{$<$}\kern-.85em
          \lower.35em\hbox{$\sim$}~}\xspace}

\def\sPlot{\mbox{\em sPlot}\xspace}

\def\pt         {\mbox{$p_{\mathrm{ T}}$}\xspace}

\def\evtgen     {\mbox{\textsc{EvtGen}}\xspace}

\def\geant      {\mbox{\textsc{Geant4}}\xspace}

\def\photos     {\mbox{\textsc{Photos}}\xspace}

\def\pythia     {\mbox{\textsc{Pythia}}\xspace}

\def\tell1  {TELL1\xspace}
\def\ukl1   {UKL1\xspace}

\usepackage{cite} 
\usepackage{mciteplus}
\usepackage{mathrsfs}

\usepackage{longtable} 

\begin{document}

\renewcommand{\thefootnote}{\fnsymbol{footnote}}
\setcounter{footnote}{1}

\begin{titlepage}
\pagenumbering{roman}

\vspace*{-1.5cm}
\centerline{\large EUROPEAN ORGANIZATION FOR NUCLEAR RESEARCH (CERN)}
\vspace*{1.5cm}
\noindent
\begin{tabular*}{\linewidth}{lc@{\extracolsep{\fill}}r@{\extracolsep{0pt}}}
\ifthenelse{\boolean{pdflatex}}
{\vspace*{-1.5cm}\mbox{\!\!\!\includegraphics[width=.14\textwidth]{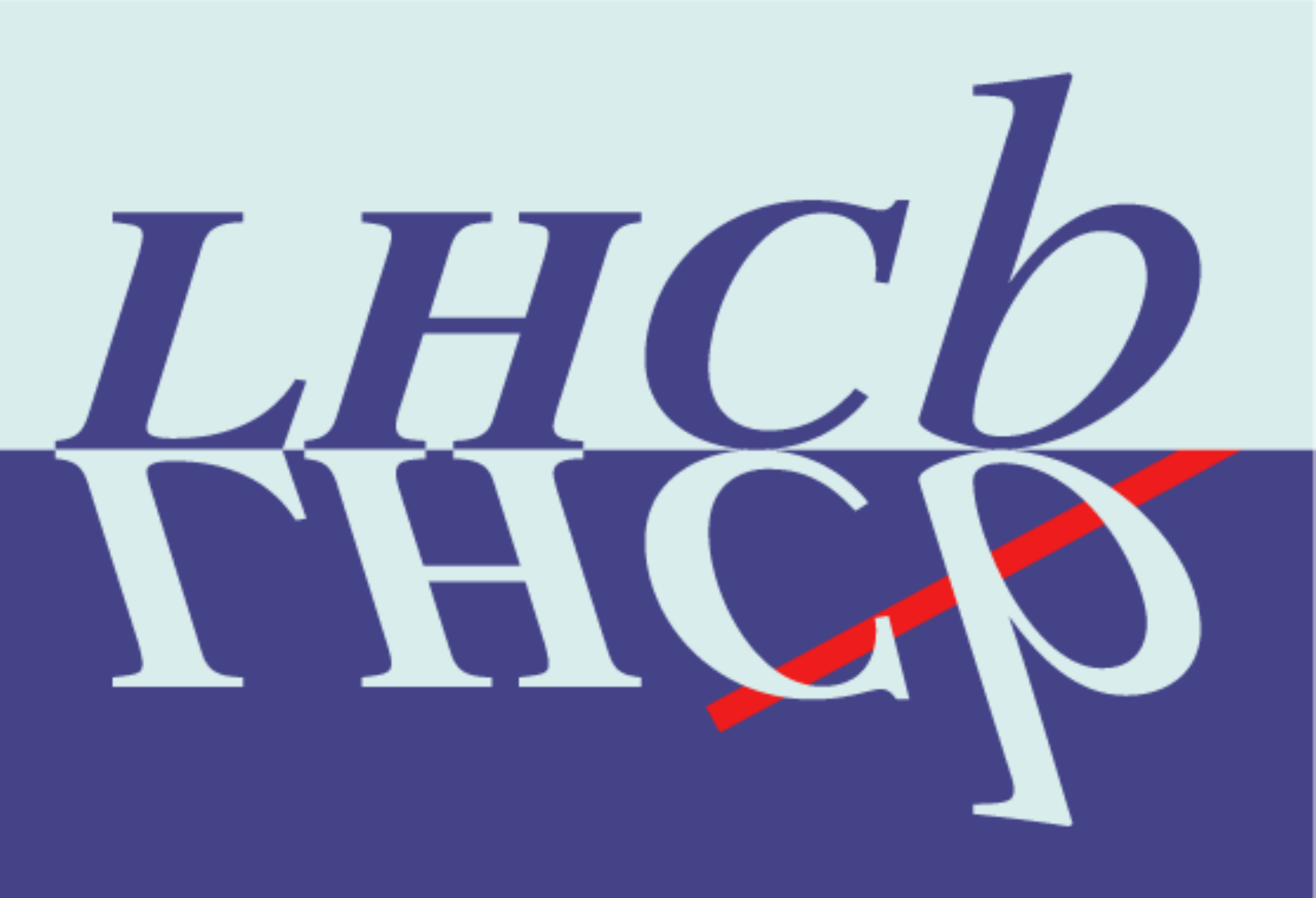}} & &}%
{\vspace*{-1.2cm}\mbox{\!\!\!\includegraphics[width=.12\textwidth]{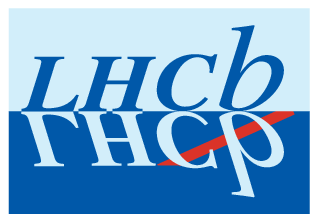}} & &}%
\\
 & & CERN-EP-2018-051 \\  
 & & LHCb-PAPER-2018-005 \\  
 & & April 25, 2018\\ 
 & & \\
\end{tabular*}

\vspace*{2.0cm}

{\normalfont\bfseries\boldmath\huge
\begin{center}
  \papertitle
\end{center}
}

\vspace*{1.0cm}

\begin{center}
\paperauthors\footnote{Authors are listed at the end of this Letter.}
\end{center}

\begin{abstract}
  \noindent
  The decay $\Lb \to \Lc  p \overline{p} \pi^-$ is observed using $pp$ collision 
  data collected with the LHCb detector at centre-of-mass energies of $\sqrt{s}=$
  7 and 8\tev, corresponding to an integrated luminosity of
  3\invfb. The ratio of branching fractions between $\Lb \to \Lc  p \overline{p} \pi^-$
  and $\Lb \to \Lc \pi^-$ decays is measured to be
  \begin{equation*}
      \frac{\mathcal{B}(\Lb \to \Lc  p \overline{p}\pi^-)}{\mathcal{B}(\Lb \to \Lc \pi^-)} = 0.0540 \pm 0.0023 \pm 0.0032.
  \end{equation*}
  Two resonant structures are observed in the $ \Lc \pi^-$ mass spectrum of the
  ${\Lb \to  \Lc  p\overline{p} \pi^-}$ decays, corresponding to the
  $\PSigma_c(2455)^0$ and $\PSigma_c^{*}(2520)^0$ states. The ratios of branching
  fractions with respect to the decay $\Lb \to \Lc  p \overline{p} \pi^-$
  are
  \begin{align*}
    \frac{\mathcal{B}(\Lb \to \PSigma_c^0 p\overline{p})\times\mathcal{B}(\PSigma_c^0\to \Lc \pi^-)}{\mathcal{B}(\Lb \to \Lc  p \overline{p}\pi^-)} = 0.089\pm0.015\pm0.006,\\
    \frac{\mathcal{B}(\Lb \to \PSigma_c^{*0} p\overline{p})\times\mathcal{B}(\PSigma_c^{*0}\to \Lc \pi^-)}{\mathcal{B}(\Lb \to \Lc  p \overline{p}\pi^-)} = 0.119\pm0.020\pm0.014.
  \end{align*}
In all of the above results, the first uncertainty is statistical and the second is systematic. 
The phase space is also examined for the presence of dibaryon resonances.
No evidence for such resonances is found.
\end{abstract}

\vspace*{2.0cm}

\begin{center}
   Published in Phys.~Lett.~B 784 (2018) 101-111
\end{center}

\vspace{\fill}

{\footnotesize
\centerline{\copyright~\papercopyright. \href{\paperlicenceurl}{\paperlicence}.}}
\vspace*{2mm}

\end{titlepage}

\newpage
\setcounter{page}{2}
\mbox{~}

\cleardoublepage

\renewcommand{\thefootnote}{\arabic{footnote}}
\setcounter{footnote}{0}

\pagestyle{plain} 
\setcounter{page}{1}
\pagenumbering{arabic}


\section{Introduction}
\label{sec:Introduction}

The quark model of Gell-Mann~\cite{GellMann:1964nj} and Zweig~\cite{Zweig:1964} classifies mesons 
(\qqbar) and baryons (\quark\quark\quark) into multiplets, and also allows for hadrons 
with more than the minimal quark contents.
In 2015, LHCb observed two pentaquark states in the decay
${\Lb\to\jpsi \proton \Km}$\cite{LHCb-PAPER-2015-029}. In the decay  
channel  
${\Lb\to\Lc\proton\antiproton\pim}$,\footnote{Unless explicitly noted, charge conjugate decays are implied.}  
charmed dibaryon resonant states 
could be present. As discussed 
in Ref.~\cite{MAIANI201537}, such states could manifest via the decay
${\Lb \to \antiproton + [cd][ud][ud] = \antiproton + \mathscr{D}_c^+}$, where
${\mathscr{D}_c^+}$ is the dibaryon state with a mass below 4682\mevcc.
The subsequent decay of the ${\mathscr{D}_c^+}$ dibaryon could proceed either via
quark rearrangement to the final state ${\proton\PSigma_c^0}$, with
${\PSigma_c^0\to \Lc\pim}$, or via string breaking to the final state $\mathscr{P}_c^0(\bar{u}[cd][ud])$, 
which could involve
a lighter, yet undiscovered $\mathscr{P}_c^0$  pentaquark state,  
$\mathscr{D}_c^+\to \mathscr{P}_c^0(\bar{u}[cd][ud]) \proton$,  
with ${\mathscr{P}_c^0 \to \Lc\pim}$~\cite{MAIANI201537}.
The discovery of any of these decay modes would test the predictions 
of quantum chromodynamics and the fundamental workings of the Standard Model.

In this Letter, the first observation of the decay ${\Lb\to\Lc\proton\antiproton\pim}$, 
referred to as the signal channel, is reported. 
A measurement is made of its  
branching fraction relative to the normalisation channel ${\Lb\to\Lc\pim}$.  
Resonance structures within the ${\Lc\proton\antiproton\pim}$ system are also investigated. 
While no evidence for dibaryon resonances is found, 
significant contributions from the $\PSigma_c(2455)^0$ and $\PSigma_c^{*}(2520)^0$ resonances 
are found in the ${\Lc\pim}$ invariant mass spectrum. 
The ratios of branching fractions between decays via these resonances, 
hereinafter denoted as $\PSigma_c^0$ and $\PSigma_c^{*0}$, and the 
${\Lc\proton\antiproton\pim}$ inclusive decay are also reported.
The measurements in this Letter  are based on a data sample of $pp$ collisions collected  
with the LHCb detector at centre-of-mass energies of $\sqrt{s}=$
7\tev in 2011 and $\sqrt{s}=$ 8\tev in 2012, corresponding to an integrated luminosity of
3\invfb.

\section{Detector and simulation}
\label{sec:Detector}

The \lhcb detector~\cite{Alves:2008zz,LHCb-DP-2014-002} is a single-arm forward
spectrometer covering the \mbox{pseudorapidity} range ${2<\eta <5}$,
designed for the study of particles containing \bquark or \cquark
quarks. The detector includes a high-precision tracking system
consisting of a silicon-strip vertex detector surrounding the $pp$
interaction region, a large-area silicon-strip detector located
upstream of a dipole magnet with a bending power of about
$4{\mathrm{\,Tm}}$, and three stations of silicon-strip detectors and straw
drift tubes placed downstream of the magnet.
Different types of charged hadrons are distinguished using information
from two ring-imaging Cherenkov (RICH) detectors.
Photons, electrons and hadrons are identified by a calorimeter system consisting of
scintillating-pad and preshower detectors, an electromagnetic
calorimeter and a hadronic calorimeter. Muons are identified by a
system composed of alternating layers of iron and multiwire
proportional chambers.
The online event selection is performed by a trigger~\cite{LHCb-DP-2012-004},
which consists of a hardware stage, based on information from the calorimeter and muon
systems, followed by a software stage, in which all charged particles
with ${\pt>500\,(300)\mevc}$ are reconstructed for 2011\,(2012) data, where \pt is the transverse momentum~\cite{LHCb-DP-2012-004}.  
At the hardware trigger stage, events are required to contain a 
muon or dimuon pair with high \pt, or a hadron, photon or electron 
with high transverse energy deposited in the calorimeters.
The software trigger requires a two-, three- or four-track
secondary vertex with a significant displacement from 
any primary proton-proton interaction vertices (PVs).  
At least one charged particle
must have a ${\pt > 1.7~(1.6)\gevc}$ for 2011 (2012) data, and be
inconsistent with originating from a PV.
A multivariate algorithm~\cite{BBDT} is used for
the identification of secondary vertices consistent with the decay
of a \bquark hadron.

Simulated samples of the signal, the normalisation channels and 
backgrounds produced in $pp$ collisions  are generated using
\pythia~\cite{Sjostrand:2007gs,*Sjostrand:2006za}
with a specific \lhcb
configuration~\cite{LHCb-PROC-2010-056}. Decays of hadronic particles
are described by \evtgen~\cite{Lange:2001uf}, in which final-state
radiation is generated using \photos~\cite{Golonka:2005pn}. The
interaction of the generated particles with the detector, and its response,
are implemented using the \geant
toolkit~\cite{Allison:2006ve, *Agostinelli:2002hh} as described in
Ref.~\cite{LHCb-PROC-2011-006}.

\section{Candidate selection}
The ${\Lb\to\Lc\proton\antiproton\pim}$ and ${\Lb\to\Lc\pim}$ candidates 
are reconstructed using the decay ${\Lc\to\proton\Km\pip}$. 
An offline selection is applied,  
based on a loose preselection, followed by a multivariate analysis. 
To minimize the systematic uncertainty on the ratio of efficiencies  
between the signal and the normalisation channels, the selection criteria on the \Lc candidates   
are similar  between the two channels.

Reconstructed final-state particles in ${\Lb\to\Lc\proton\antiproton\pim}$ and 
${\Lb\to\Lc\pim}$ candidate decays are required to have 
a momentum ${p>1\gevc}$ and ${\pt>100\mevc}$.  
Protons and antiprotons are required to 
have ${p>10\gevc}$ to improve particle identification.  
All final-state particles are also required to be inconsistent with originating from any PV, 
by rejecting the tracks with a small \chisqip, where 
\chisqip is the difference in the vertex-fit \chisq of a given PV 
with or without the track considered, requiring ${\chisqip > 4}$. 
Candidate \Lc decays are required to have at least one decay product with ${\pt>500\mevc}$ and 
${p>5\gevc}$, a good vertex-fit quality, and an invariant mass 
within $\pm$15\mevcc of the known \Lc mass~\cite{PDG2016}. The scalar sum of the transverse momenta of 
the \Lc decay products is required to be greater than 1.8\gevc. 

The \Lc\pim candidate is reconstructed by combining a \Lc candidate 
with a pion, and the signal candidate is reconstructed by combining a \Lc candidate 
with a pion, a proton and an antiproton.  
These combinations must form a \Lb candidate with a good-quality vertex and  
be consistent with 
originating from the associated PV,  
defined as that for which the \Lb candidate has the least \chisqip. 
Furthermore, the \Lc candidate is required to decay downstream of the \Lb 
decay vertex. The \Lb decay time, calculated as ${t = m_\Lb L/p}$, is required to be 
greater than 0.2\ps, where $m_\Lb$ is the mass, $L$ is the decay length and $p$ is 
the momentum of the $\Lb$ candidate.
The \Lb candidate is also required to have at least one final-state particle in the decay chain with 
${\pt>1.7\gevc}$, ${p>10\gevc}$, and have at least one track significantly 
inconsistent with originating from the associated PV by requiring the track to have  ${\chisqip > 16}$. 
Final-state tracks of signal and normalisation channel candidates  
must pass strict particle-identification requirements based on the RICH
detectors, calorimeters and muon stations. 
A constrained fit~\cite{Hulsbergen:2005pu} is applied to the candidate decay chain for 
both the signal and the normalisation channels,  
requiring the \Lb candidate to come from the associated PV and constraining the \Lc particle to its known 
mass~\cite{PDG2016}. In the case of the search of the resonant  contributions, the mass of the 
\Lb candidate is also constrained to the known mass~\cite{PDG2016}. 

Trigger signals are associated with reconstructed
particles from the decays of the signal channel or of the normalisation channel. 
Selection requirements can therefore be made on the trigger
selection itself and on whether the decision was due to the reconstructed candidate decay, 
other particles produced in the $pp$ collision, or a combination of the two.
This association makes it possible to use a data-driven method for the correction
and systematic uncertainty estimation on the trigger efficiencies~\cite{LHCb-DP-2012-004}.
To take advantage of the similarity between the signal and the normalisation channels,  
which helps to  minimize the systematic uncertainty on the ratio of their efficiencies,
candidates are classified in one of the following two hardware trigger categories.
In the first category, called Triggered On Signal ({TOS}), the candidate must include a hadron consistent with originating from
the decay of a $\Lc$ candidate and which deposited enough transverse energy  in the calorimeter
to satisfy the hardware trigger requirements. The typical value of the transverse energy
threshold is around 3.5\gevcc.
As the \Lc baryon is a \Lb decay product for both the signal  and the 
normalisation channels, this choice minimizes the difference between the \Lb decay modes. 
The second category,
called Triggered Independent of Signal ({TIS}), comprises events
which satisfied the hardware trigger through signatures unassociated with the complete \Lb decay chains, 
either due to a muon with high \pt,
or a hadron, photon, or electron with high transverse
energy deposited in the calorimeters.
The efficiencies of the TIS and TOS requirements are different, so
the data are divided into two
statistically independent samples, one TIS, and the other TOS and not TIS,
which will be referred to as TOS for the rest of this Letter.

The so-called cross-feed backgrounds, contributing under the peak of the 
invariant mass of the normalisation channel or of the signal channel  
from the  ${\Bzb(\Bsb)\to \Dp(\Dsp) \pim}$ and  
${\Bzb(\Bsb) \to \Dp(\Dsp) \proton \antiproton \pim}$ decays, respectively, with ${\Dp(\Dsp) \to\Kp\Km\pip}$ 
or ${\Dp \to \Km\pip\pip}$,  
where either the kaon or pion is misidentified as a proton,  
are explicitly vetoed when both of the following two conditions are satisfied. First, the mass hypothesis 
of the proton from the \Lc candidate is replaced with either the kaon or pion hypothesis,  
and the resulting invariant mass of the combination is consistent  
with the known $\Dp(\Dsp)$ mass~\cite{PDG2016} within $\pm$15\mevcc.
Second, the invariant mass of the \Lc candidate 
is set to the known $\Dp(\Dsp)$ mass~\cite{PDG2016}, and the resulting invariant mass of the \Lb candidate is consistent with  
the known $\Bzb(\Bsb)$ mass~\cite{PDG2016} within $\pm$25\mevcc for ${\Lb\to\Lc\proton\antiproton\pim}$ decays, and within  
$\pm$45\mevcc for ${\Lb\to\Lc\pim}$ decays.

Further background reduction is achieved using a multivariate
analysis based on a gradient boosted decision tree (BDTG)\cite{Breiman}. The BDTG
is trained using twelve variables: the vertex-fit quality of the \Lc 
and \Lb candidates, the decay-vertex displacement along the beamline between the \Lb and \Lc candidates, 
the displacement between the decay vertex of the \Lb candidate and the associated PV, the 
\chisqip of the \Lb candidate, the angle between the reconstructed \Lb momentum and 
the direction of flight from the associated PV to the decay vertex, 
the smallest \pt  
and smallest \chisqip among the three \Lc decay products, the \pt and \chisqip of the pion originating directly  
from the \Lb decay, and the smallest \pt and smallest \chisqip between  
the \proton and \antiproton originating directly from the \Lb decay.   
The BDTG training is performed using simulated samples for the signal,  
and data distributions for the background, 
with reconstructed invariant mass 
well above the known \Lb mass~\cite{PDG2016}. Cross-feed backgrounds from the
decays ${\Lb\to\Lc\Kp\Km\pim}$, ${\Bzb\to\Lc\antiproton\pip\pim}$ and ${\Bsb\to\Lc\antiproton\Kp\pim}$
are explicitly vetoed during the BDTG-training process by requiring the difference between the    
reconstructed \bquark-hadron mass and its known mass to be  
greater than $\pm$30\mevcc. The BDTG selection is optimized for the figure of merit  
$S/\sqrt{S+B}$, where $S$ and $B$ are the expected signal and background yields  
within $\pm$30\mevcc of the known \Lb mass~\cite{PDG2016}. The initial value of $S$ and $B$  
without BDTG selection is obtained from the \Lb mass spectrum in data.   
No improvement in the normalisation channel
is found using a similar procedure, therefore no BDTG selection is applied. 
A systematic uncertainty is assessed for this choice in Section~\ref{sec:syst}.   

Due to the large number of final-state particles in the \Lb decays, particles   
with the same charge may share track segments, representing a possible background. 
These  tracks are referred to as clones, 
and are suppressed by requiring    
that the opening angle between any same-charged tracks in the candidate is larger than 0.5 mrad. 
This selection
removes 2\% of candidates in the signal sample and 0.1\% in the normalisation sample. 
If multiple \Lb candidates are reconstructed in one single event, 
one candidate is chosen at random in the following two cases. 
First, if the proton from the \Lc decays is exchanged with that directly from 
the \Lb decays, forming two candidates with nearly the same \Lb mass. 
Second, if a track from one candidate shares a segment with a track from another  
candidate. 
With these criteria, 2.5\% of candidates 
in the signal sample and 0.1\% in the normalisation sample are vetoed. 
After these selections, 0.8\% of events in the signal sample  
and 0.2\% in the normalisation sample contain multiple \Lb candidates. 
These remaining multiple candidates mainly originate from the random 
combinations of the final-state tracks, and have a negligible influence on the 
estimation of the signal yields.  
No further vetoes on these candidates are applied.

\section{Efficiencies}
\label{sec:eff}
The total efficiencies of the signal and the normalisation decays are given by
\begin{equation}
\epsilon_{\rm total} = \epsilon_{\rm a}\cdot\epsilon_{\rm rec\&sel|a}\cdot\epsilon_{\rm trig|sel}\cdot\epsilon_{\rm PID},
\end{equation}
where $\epsilon_{\rm a}$ represents the geometrical acceptance of the \lhcb detector,
$\epsilon_{\rm rec\&sel|a}$ is the efficiency of reconstruction and selection  
calculated on candidates in the acceptance, $\epsilon_{\rm trig|sel}$ is the trigger efficiency  
of the selected candidates, and $\epsilon_{\rm PID}$ is the particle-identification efficiency.
All efficiencies except $\epsilon_{\rm PID}$ and $\epsilon_{\rm trig|sel}$
are determined from simulation. The particle-identification efficiency is determined  
from calibration data specific to each data-taking year, binned in momentum and pseudorapidity
of the track in question, as well as in the multiplicity of the event~\cite{LHCb-PROC-2011-008}.
The trigger efficiency is determined from a combination of
simulation and data-driven techniques where the agreement between data and simulation
is explicitly verified using the normalisation sample satisfying the TIS requirement. 
All efficiencies are calculated separately for the TIS and TOS
trigger samples, and for data-taking year, due to the difference in
centre-of-mass energies. Agreement between data and simulation is improved by applying
a per-candidate weight to the \pt and rapidity, $y$, of the \Lb baryon in simulated events to match
the normalisation sample in the TIS category, which is largely independent of trigger
conditions. 
The \pt and $y$ distributions of \Lb produced in $pp$ collision 
are identical for the signal and the normalisation channels, so the same per-candidate weights 
are applied to the signal sample. 
The simulated \chisqip of the final-state particles and the vertex-fit \chisq of \Lc candidates  
are weighted to reproduce the data distributions. 
The ratio between the efficiencies of the signal and the normalisation channels,  $\epsilon_r$, 
is $(10.00\pm0.12)\%$ for the TIS sample and $(11.39\pm0.22)\%$ for the
TOS sample, including uncertainties due to the limited size of the simulated sample.

\section{Fit model and the ratio of branching fractions}
\label{sec:eff}
\begin{figure}[!tbp]
\begin{center}
\includegraphics[width=0.48\textwidth]{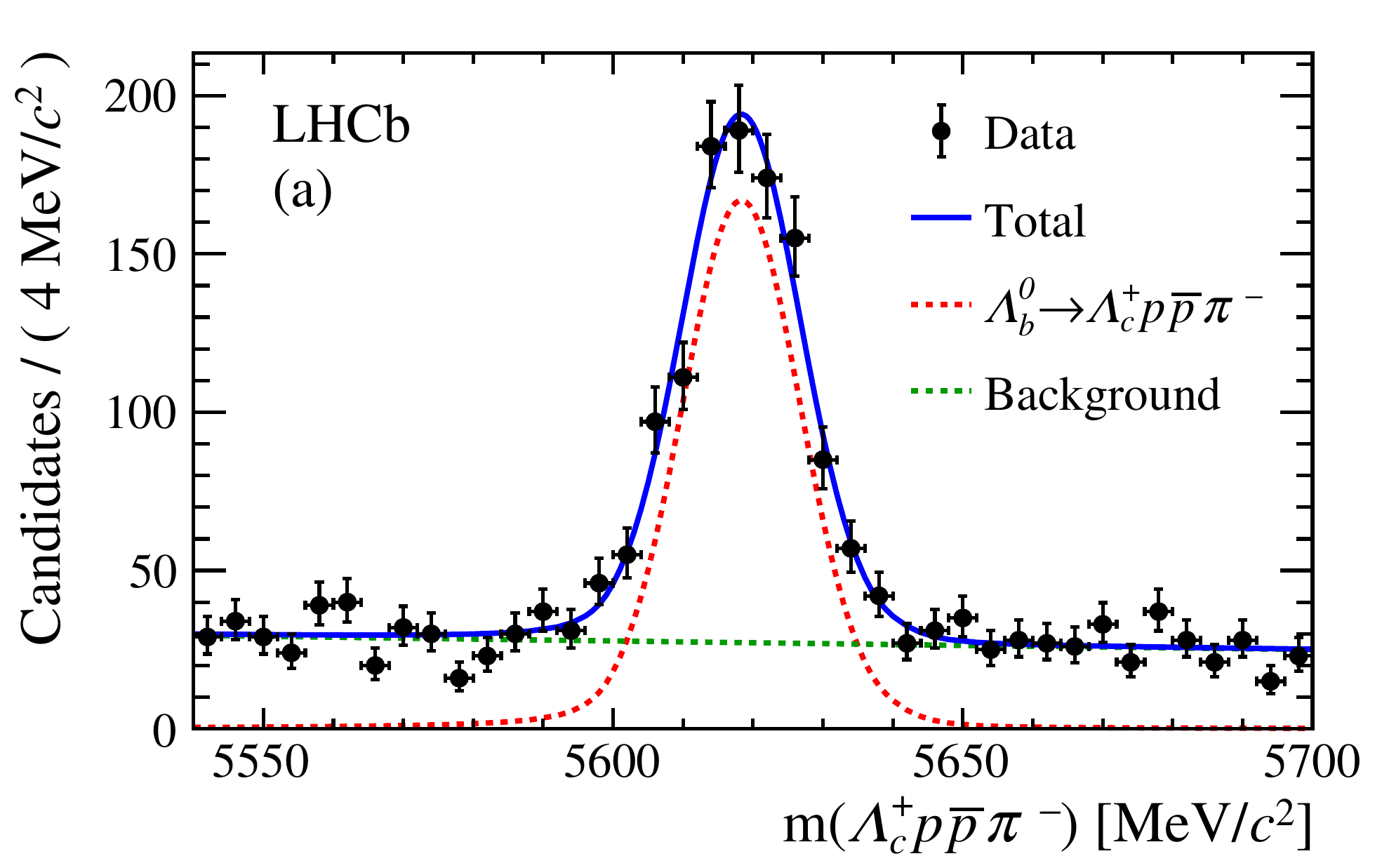}
\includegraphics[width=0.48\textwidth]{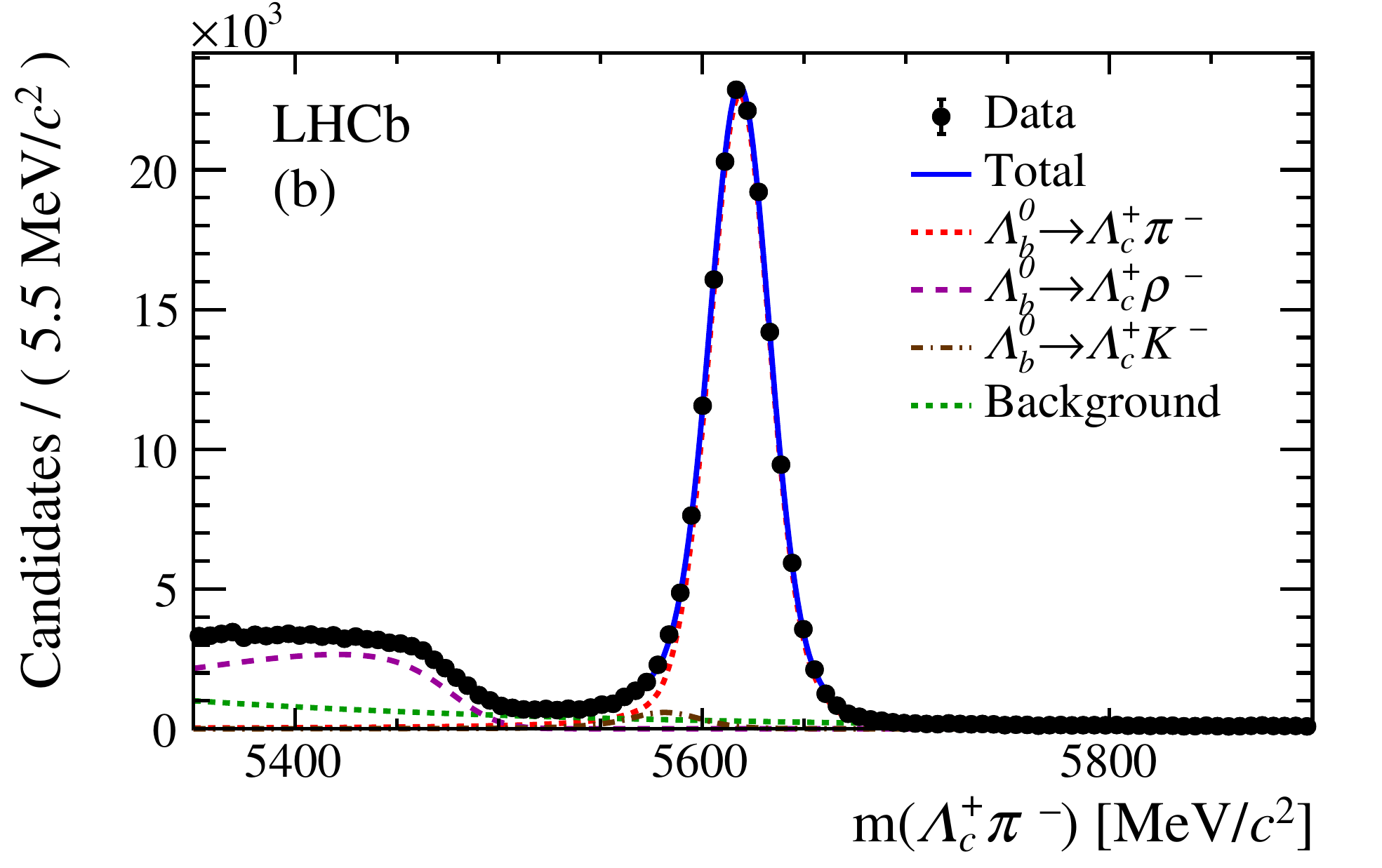}
\caption{Invariant mass distributions of the (a) $\Lb\to\Lc\proton\antiproton\pim$ 
and (b) $\Lb\to\Lc\pim$ candidates. Fit results are overlaid as a solid blue line.
For (a), the red dotted line represents the signal component and the green dotted
line the background due to random combinations. For (b), the red dotted line is the
signal component, the green dotted line is the random combination background,
the purple dashed line is the contribution from $\Lb\to\Lc\rhom$ and the brown
dashed-dotted line represents the contribution from $\Lb\to\Lc\Km$.}
\label{fig:fits}
\end{center}
\end{figure}

The yields in both the signal and  
the normalisation channels are determined  
from an unbinned extended maximum-likelihood fit to the corresponding invariant-mass spectra 
with both the TIS and TOS samples combined.  
The signal  
is modelled by a sum of two Crystal Ball functions\cite{Skwarnicki:1986xj} with
a common mean of the Gaussian core, and with the tail parameters fixed from simulation. For both the
signal and the normalisation channels, the background from random combinations of final-state  
particles is described by an exponential function, whose parameters are left free in the fits 
and are independent between the signal and the normalisation channels.  
For the
normalisation channel, background from the $\Lb\to\Lc\rhom$ decays, with
$\rhom\to\pim\piz$ is modelled by the convolution of an empirical threshold function  
with a Gaussian resolution. The contribution due to misidentification of the kaon to pion from $\Lb\to\Lc\Km$
is modelled by a sum of two Crystal Ball functions. The parameters of these two background sources are
taken from simulation. The fits to the invariant-mass distributions for the signal 
and the normalisation channels are shown in Figure~\ref{fig:fits}. In this figure, the TIS and TOS 
samples are combined. From these fits, $926\pm43$
$\Lb\to\Lc\proton\antiproton\pim$ and $(167.00\pm0.50)\times10^3$   
$\Lb\to\Lc\pim$ decays are observed.

To determine the ratio of  branching fractions  
${\frac{\mathcal{B}(\Lb \to \Lc  p \overline{p}\pi^-)}{\mathcal{B}(\Lb \to \Lc \pi^-)} }$, 
indicated  in the  following by $\mathcal{B}_r$,  
a simultaneous fit is performed to the signal and the normalisation channels, each divided into the two  
independent trigger categories. The yield of the normalisation sample, $N(\Lb\to\Lc\pim)$, is 
a free parameter in the fits, whereas the yield of the signal sample is calculated as
${N(\Lb\to\Lc\proton\antiproton\pim) = \mathcal{B}_r\times\epsilon_r\times N(\Lb\to\Lc\pim)}$, where
$\epsilon_r$ is the ratio between the total efficiency of the ${\Lb\to\Lc\proton\antiproton\pim}$ and ${\Lb\to\Lc\pim}$ decays.  
The ratio of branching fractions $\mathcal{B}_r$  
is the same for the TIS and TOS subsamples and is measured to be  
${\mathcal{B}_r = 0.0542 \pm 0.0023}$. The corresponding
signal yields are 677 $\pm$ 29 for the TIS subsample and 259 $\pm$ 11 for the TOS subsample; the yields in the normalisation
sample are ${(124.9 \pm 0.4)\times 10^3}$ for the TIS subsample and ${(41.9 \pm 0.2) \times10^3}$ for the TOS subsample.

\section{Systematic uncertainties}
\label{sec:syst}
The systematic uncertainties on the measurement of the ratio of branching 
fractions are listed in Table~\ref{tab:systemerr}. The total systematic uncertainty  
is determined from the sum in quadrature of all terms. 

First, the uncertainty
related to the background modelling is considered.
In the signal sample, 
the exponential function is replaced with a second-order polynomial for the background component. 
For the normalisation
channel, the model is varied by using the sum of
two exponential functions. The resulting uncertainty on the ratio of branching  
fractions is 0.6\%. The uncertainties due to the ${\Lb\to\Lc\Km}$ shape parameters are assessed
by increasing the width of the Crystal Ball functions  
by 10\%, corresponding to two standard deviations, resulting in a change of 0.1\%.
The uncertainty due to the ${\Lb\to\Lc\rhom}$ contribution is estimated by varying
the shape parameters by one standard deviation, resulting in an uncertainty of 0.4\%.
The total uncertainty on the ratio of the branching fractions due to the background modelling 
is 0.7\%.  

The signal-model parameterization is changed to a single Hypatia function\cite{Santos:2013gra},
where the mean and width are allowed to float and all other parameters are taken  
from simulation, resulting in an uncertainty of 0.1\%.

\begin{table}[!tbp]
\centering
\caption{Summary of systematic uncertainties and correction factors
to the ratio of branching fractions measurement. 
All uncertainties 
are given as a percentage of the ratio  
of branching fractions.}
\label{tab:systemerr}
\resizebox{\textwidth}{!}{
\begin{tabular}{lSS}\hline\hline
\multicolumn{1}{l}{Source} & \multicolumn{1}{c}{Uncertainty (\%)}  & \multicolumn{1}{c}{Correction factor} \\\hline
Background fit model & 0.7 & $\rm{-}$ \\
Signal fit model & 0.1 & $\rm{-}$ \\
PID efficiency &0.3 & $\rm{-}$ \\
Tracking efficiency calibration &0.8 & 0.985\\
Kinematic range of final-state tracks&0.7 & $\rm{-}$ \\
Hadron interaction &4.4 & $\rm{-}$ \\
$\pt, ~y$ weighting  & 1.0 & $\rm{-}$ \\
Trigger efficiency  &2.9 & $\rm{-}$\\
Simulated sample size  &1.3 & $\rm{-}$\\
Candidates with clone tracks and multiple candidates & 0.2&$\rm{-}$ \\
Veto of the reflection background & 0.4 & $\rm{-}$ \\
\Lc Dalitz weighting &0.2  &0.984\\
\Lc polarization & 0.3  & 0.987\\
Resonant structures & 1.8& 1.041\\
\hline
Total&6.0 &0.996 \\\hline
\end{tabular}
}
\end{table}

The uncertainty on the relative efficiency of the particle identification is assessed by
generating pseudoexperiments. For each pseudoexperiment, efficiencies in different 
momentum, pseudorapidity and multiplicity bins are  
determined from independent Gaussian distributions with mean values equal to the nominal efficiencies and widths 
corresponding to their uncertainties. 
This procedure is repeated 1000 times, and
the width of the resulting efficiency is taken as the
systematic uncertainty. This procedure, performed separately for the TIS and TOS
samples, results in a 0.13\% uncertainty for both samples. Binning effects on 
the efficiency are estimated by halving the bin size of the momentum distributions, resulting
in a relative change of 0.2\% for the TIS sample and 0.1\% for the TOS sample.
The total uncertainty on the relative efficiency for the TIS and TOS samples is   
0.24\% and 0.16\%, respectively, 
corresponding to an uncertainty of 0.3\% on the ratio of the branching fractions. 

Tracking efficiencies are determined with simulated
events weighted to match the kinematic properties of dedicated calibration samples.
The weights are determined  
as a function of the kinematic variables,  
separately for each data-taking year~\cite{LHCb-DP-2013-002}. 
The kinematic properties of the \Lc
decay products are similar for
the signal and the normalisation samples and therefore provide minor    
contributions to the total tracking efficiency ratio. The dominant contribution to the systematic uncertainty  
comes from the knowledge of the \proton and \antiproton tracking efficiencies, 
whose systematic uncertainties are fully correlated. 
The efficiency correction procedure gives a change in efficiency
of 2.0\% for the TIS sample and 1.4\% for the TOS sample, yielding a total
correction factor of 0.985 for the ratio of branching fractions, and 
a systematic uncertainty of 0.4\% for each of the \proton and \antiproton 
 mainly stemming from the finite size of the calibration sample~\cite{LHCb-DP-2013-002}. 

Due to distinct trigger requirements, the kinematic acceptance of the
calibration samples differs slightly from the signal and the normalisation channels.
A nonnegligible  fraction of candidates have final-state  
particles in a kinematic range  outside of the regions covered 
by the calibration samples.  
About 20\% of the candidates from both channels fall in this category 
due to the low-momentum pion from the \Lc decay. In addition, 10\% 
of the candidates from the signal channel are also affected, mainly due to the pion originating from the \Lb decay. 
For all of these outside-range candidates, the efficiency correction in the nearest available 
bin is used. As the effects for \Lc decays cancel in the relative efficiency, only the additional 
10\% candidates in the signal channel contribute a 0.7\% uncertainty on the relative efficiency.   

Hadronic interactions with the \lhcb
detector contribute an additional uncertainty of 2.2\% on the ratio 
of the branching fractions for each \proton or
\antiproton (4.4\% in total), which is obtained from simulation, accounting for the imperfect knowledge of  
material budget of the \lhcb detector\cite{LHCbVELOGroup:2014uea}.

Per-candidate weights depending on \pt and $y$ of the \Lb baryon are applied in 
simulated events to improve the agreements between data and simulation.  
Systematic uncertainties for the weighting due to the finite
size of the normalisation sample are assessed with pseudoexperiments.
In each pseudoexperiment, the weights are varied within their uncertainties, 
and the results are propagated  to the ratio of branching fractions. The
standard deviation of the obtained distributions is taken as a systematic uncertainty,
resulting in 0.65\% for the TIS sample and 0.65\% for the TOS sample. 
The systematic uncertainties due to the binning scheme of the weighting in \pt and $y$ are   
estimated by halving the bin size, or using the gradient boosting~\cite{friedman2001greedy}\cite{Belov:2010xm}, which is an unbinned method of weighting, 
to check the changes on the relative efficiencies.  
The resulting systematic uncertainties
are 0.43\% for the TIS sample and 1.5\% for the TOS sample.  
After propagation through the entire fit procedure, this results
in an uncertainty of 1.0\% on the ratio of the branching fractions.
 
Trigger efficiencies for the TOS samples are also assessed using pseudoexperiments which are propagated
to the final measurement, resulting in a final uncertainty of 0.1\%.
The trigger efficiency of the TIS sample is taken from simulation. 
Its systematic uncertainty is computed from the difference between 
the 
TIS efficiency taken from data and simulation for events which are 
triggered both on the \Lc candidate  and also on other tracks unassociated to the signal decay.  
As a result, a systematic uncertainty of 3.9\% is assigned for the relative trigger efficiency of the TIS sample, 
corresponding to an uncertainty of 2.9\% on the ratio of the branching fractions.


The effect of the finite size of the simulated samples is assessed by considering
the possible variation of the efficiency with weighted samples in a bin of \pt and
rapidity of the \Lb candidate, and the corresponding systematic uncertainty  
on the efficiency of the signal or normalisation channel, TIS or TOS sample, is given by
\begin{equation}
\sigma_\epsilon = \sqrt{\sum_{i}\epsilon_{i}(1-\epsilon_{i})N_{i}w_{i}}/\sum_{i}N_{i}w_{i},
\end{equation}
where
for each bin $i$, $N_{i}$ is the number of candidates, $w_{i}$ is the single event weight,
and $\epsilon_{i}$ is the single event efficiency. 
The total uncertainty on the relative efficiency for the TIS and TOS samples is  
1.2\% and 1.9\%, respectively,
corresponding to an uncertainty of 1.3\% on the ratio of the branching fractions.

The uncertainty due to the removal of candidates reconstructed 
with clone tracks and multiple candidates is assessed by applying
the same procedure to simulation, resulting in a difference of 0.2\%.

Vetoes on the invariant mass of possible cross-feed backgrounds may bias the signal mass distributions.
An uncertainty of 0.4\% is determined 
by changing the fit range of the normalisation
sample to begin at 5450\mevcc, instead of 5350\mevcc.  

The agreement between data and simulation in the ${\Lc\to\proton\Km\pip}$ decay
is also tested by comparing the Dalitz plot distributions. The normalisation sample  is
weighted in the $m^2(\proton\Km)$ versus $m^2(\Km\pip)$ plane. Due to the smaller
sample size of the signal channel, weights obtained from the normalisation channel are
applied to the signal. The resulting procedure renders all distributions consistent
within one statistical standard deviation. The difference in the ratio of  branching
fractions is 1.3\% smaller than the nominal result, providing a correction
factor of 0.984. 
An uncertainty of 0.2\% is determined by using an alternative binning scheme 
 and varying the Dalitz-plot weights by their statistical uncertainties. 

The polarization of the \Lb particles has been measured to be consistent with zero\cite{LHCb-PAPER-2012-057}, but   
the weak decay of the \Lb baryon may induce a polarization in the \Lc system.  
In the simulation, it is assumed that the \Lc particle is unpolarized, leading  
to a difference in angular distributions between simulation and data. A possible effect 
due to the \Lc polarization is assessed by applying
a weighting procedure to the distribution of the \Lc helicity angle,
which is defined as the angle between the \Lc flight direction in the \Lb rest frame and the
direction of the $\proton\Km$ pair in the \Lc rest frame. 
This weight is obtained through a comparison between the angular distributions in simulation 
and data for the signal and the normalisation channels individually. 
Applying this weight
to both the signal and the normalisation channels does not change the efficiency with
respect to any of the other possible angles, and leads to a change of 1.1\% in the
relative efficiency for the TOS sample and 1.4\% for the TIS sample. Propagation of these
uncertainties leads to 
a correction factor of 0.987 on the ratio of the branching fractions. 
An uncertainty of 0.3\% is determined by using an alternative binning scheme
 and varying the single-candidate weights by their statistical uncertainties.

Simulated data are generated using a phase-space model for the \Lb decay,
which does not take into
account possible resonances in the ${\Lc\proton\antiproton\pim}$ system. Upon
inspection, clear signals from the $\PSigma_c^0$ and $\PSigma_c^{*0}$
resonances are found, as described in Section~\ref{sec:sigc}.  
To assess the
effect of these resonances, the simulation is weighted to reproduce the data.
Weights are applied in two invariant mass dimensions, namely the $\Lc\pim$ invariant mass
and another invariant mass of any two or three body combination. 
Among these weighting strategies, applying weights in $m(\Lc\pim)$ and $m(\proton\pim)$ (option 1) leads 
to the smallest $\mathcal{B}_r$, while weights in $m(\Lc\pim)$ and $m(\proton\antiproton\pim)$ (option 2) leads to the 
largest $\mathcal{B}_r$. 
A correction factor is computed as the average of the central values of the ratio of  branching fractions 
for the two options divided by the nominal branching fraction, with an 
uncertainty determined by half the difference between the two ratios of branching fractions.
This leads to a correction factor of 1.041 and a resulting systematic uncertainty of
1.8\%.

Uncertainties due to the use of the BDTG are tested by repeating the BDTG  training  
and selection procedure to the normalisation channel 
without variables related to the \proton \antiproton pair; 
the ratio of branching fractions is found to be consistent.

\section{\boldmath Resonance structures in the $\Lc\pim$ mass spectrum}
\label{sec:sigc}

As the resonant structure of ${\Lb\to\Lc\proton\antiproton\pim}$ decays is  
unexplored, the 
resonances in the $\Lc\pim$ system are studied.
An unbinned maximum-likelihood fit of the $\Lc\pim$ mass is performed 
for those candidates which pass all the selection criteria for the signal
${\Lb\to\Lc\proton\antiproton\pim}$ decays, to determine if there are 
resonant contributions. In this case the \Lb candidate is constrained to its known
mass~\cite{PDG2016} when obtaining the $\Lc\pim$ invariant mass spectrum.

\begin{figure}[!tbp]
\begin{center}
\includegraphics[width=0.7\textwidth]{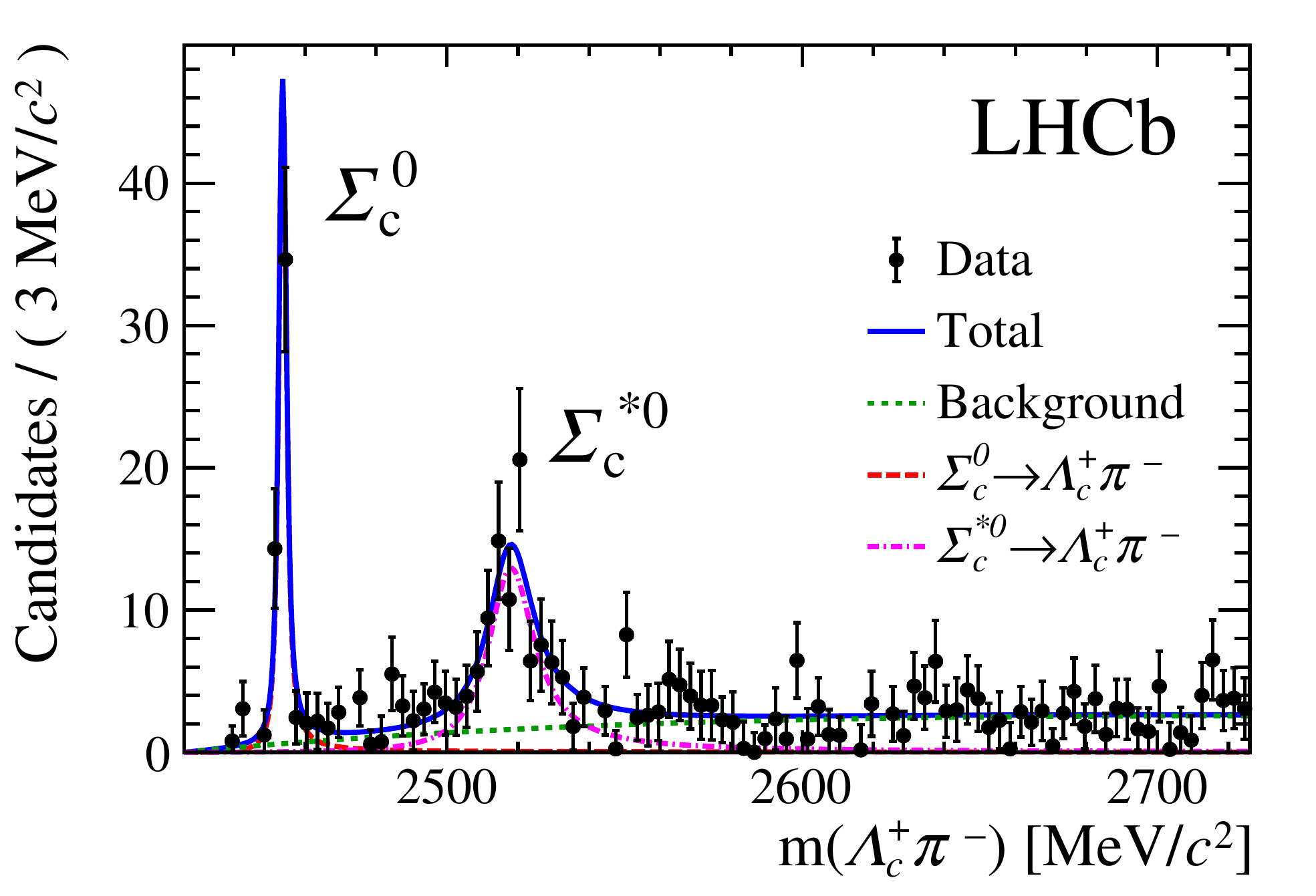}
\caption{Invariant mass of the \Lc\pim system from the decay ${\Lb\to\Lc\proton\antiproton\pim}$.
The $\PSigma_c^0$ and $\PSigma_c^{*0}$ resonances are indicated. The fit to the
data is shown as a blue continuous line, with the background component shown as a green dotted line, 
the $\PSigma_c^{0}$ shape shown as a dashed red line, and 
 the $\PSigma_c^{*0}$ shape shown as a dash-dotted magenta line.}
\label{fig:sigcres}
\end{center}
\end{figure}

The signal shapes of the $\PSigma_c^0$ 
and $\PSigma_c^{*0}$ resonances are given as the modulus squared of the relativistic 
Breit-Wigner function\cite{PDG2016}, 
\begin{equation}
\left|{\rm BW}(m|M_{0},\Gamma_{0})\right|^2 = \left|1/(M_{0}^2-m^2 -iM_{0}\Gamma(m))\right|^2, 
\end{equation}
multiplied by $m \Gamma(m)$, and convolved with a Gaussian resolution determined from simulation.
Here, $M_0$ is the known value of the $\PSigma_c^0$ or $\PSigma_c^{*0}$ mass~\cite{PDG2016}, 
$m$ is the $\Lc\pim$ invariant mass, 
and $\Gamma_0$ is the mass-independent width of the resonance, namely 1.83\mevcc for the 
$\PSigma_c^0$ and 15.3\mevcc for the $\PSigma_c^{*0}$ resonance. The mass-dependent width
is given by 
\begin{equation} 
\Gamma(m) = \Gamma_0 \times \left(\frac{q}{q_0}\right)^{2L+1}\frac{M_0}{m}B_L(q,q_0,d)^2, 
\end{equation}
where $L$ is the angular momentum in the resonance decay, $q$ is the momentum of the \Lc baryon
in the $\PSigma_c^{(*)0}$ rest frame, ${q_0 \equiv q(m=M_0)}$
and $d$ stands for the size of the $\PSigma_c^{(*)0}$ particles. From parity
and angular momentum conservation, it follows that $L=1$. The width also
depends on the Blatt-Weisskopf factor $B_L(q,q_0,d)$\cite{VonHippel:1972fg}, where the value of $d$
is set to be 1\fm (5 GeV$^{-1}$ in natural units).  
The ratio of widths of the Gaussian resolution functions for the $\PSigma_c^0$ and
$\PSigma_c^{*0}$ resonances is fixed from simulation to be 1.96.
The background is described with an empirical
threshold function. The fit shown in Figure~\ref{fig:sigcres}
yields  $59\pm10$ ${\Lb\to\PSigma_c^0\proton\antiproton}$ decays and $104\pm17$
${\Lb\to\PSigma_c^{*0}\proton\antiproton}$ decays.

The relative efficiencies for the decays  ${\Lb\to\PSigma_c^0\proton\antiproton}$, with ${\PSigma_c^0\to\Lc\pim}$ 
and ${\Lb\to\PSigma_c^{*0}\proton\antiproton}$, with ${\PSigma_c^{*0}\to \Lc\pim}$  
with respect to ${\Lb\to\Lc\proton\antiproton\pim}$ decays are determined
with an analogous  procedure as that for the ${\Lb\to\Lc\proton\antiproton\pim}$ decays relative
to the ${\Lb\to\Lc\pim}$ decays, but with the trigger samples combined due to
limited sample size. The efficiencies are $0.685\pm0.021$
for the $\PSigma_c^0$ mode and $0.904\pm0.021$ for the $\PSigma_c^{*0}$ mode, relative to
${\Lb\to\Lc\proton\antiproton\pim}$.

Many of the systematic uncertainties cancel out in the measurement
of the ratio of  branching fractions, with the remaining systematic
uncertainties stemming from the yield determination. The value of $d$  
 in the Blatt-Weisskopf factor is varied between 1.5 and 0.5\fm, with
the largest variation for each resonance taken as the systematic uncertainty,
resulting in 3.4\% for the $\PSigma_c^0$ resonance and 1.9\% for the $\PSigma_c^{*0}$ resonance. The
background shape is changed to a third-order polynomial, with a relative difference
of 1.7\% for the $\PSigma_c^0$ resonance and 10.6\% for the $\PSigma_c^{*0}$ resonance taken as the systematic
uncertainty. The masses and widths of the $\PSigma_c^{(*)0}$ resonances are allowed
to float within one standard deviation of their known values~\cite{PDG2016}, resulting in a 3.8\%
difference of the raw yield for the $\PSigma_c^0$ resonance and 2.2\% difference for the
$\PSigma_c^{*0}$ resonance. All uncertainties in the relative efficiency cancel, except for
those related to the weighting due to resonant structures in the \Lc\pim system. The scaling factor of
1.041, with an uncertainty of 1.8\% on the relative efficiency, which is shown in Table~\ref{tab:systemerr}, 
is therefore used here as well. The resulting ratios of branching fractions are

\begin{align*}
  \frac{\mathcal{B}(\Lb \to \PSigma_c^0 p\overline{p})\times\mathcal{B}(\PSigma_c^0\to \Lc \pi^-)}{\mathcal{B}(\Lb \to \Lc  p \overline{p}\pi^-)} = 0.089\pm0.015\pm0.006,\\
  \frac{\mathcal{B}(\Lb \to \PSigma_c^{*0} p\overline{p})\times\mathcal{B}(\PSigma_c^{*0}\to \Lc \pi^-)}{\mathcal{B}(\Lb \to \Lc  p \overline{p}\pi^-)} = 0.119\pm0.020\pm0.014,
\end{align*}
where the first uncertainty is statistical and the second is systematic.

\section{Search for dibaryon resonances}
\begin{figure}[htbp]
\begin{center}
\includegraphics[width=0.32\textwidth]{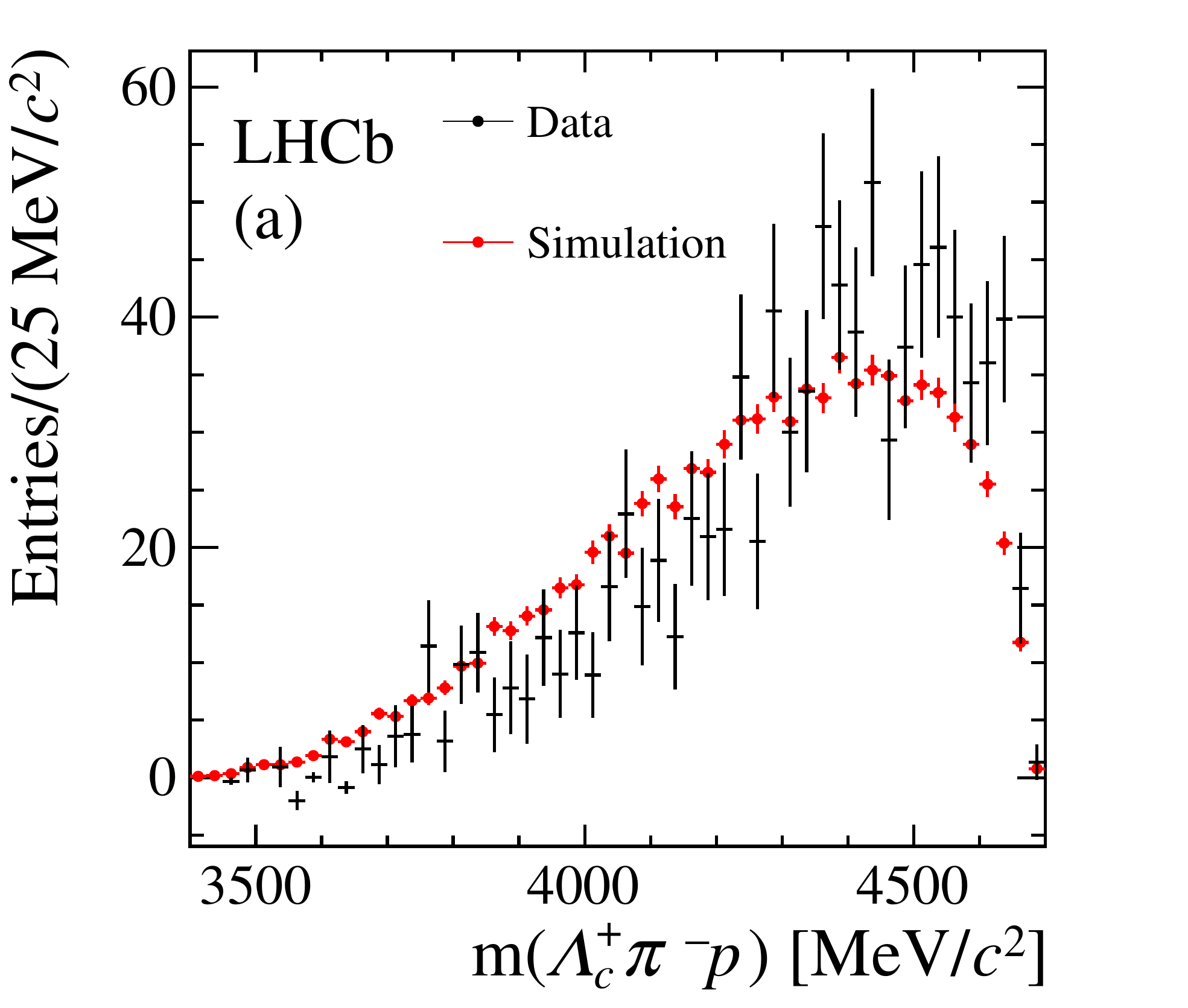}
\includegraphics[width=0.32\textwidth]{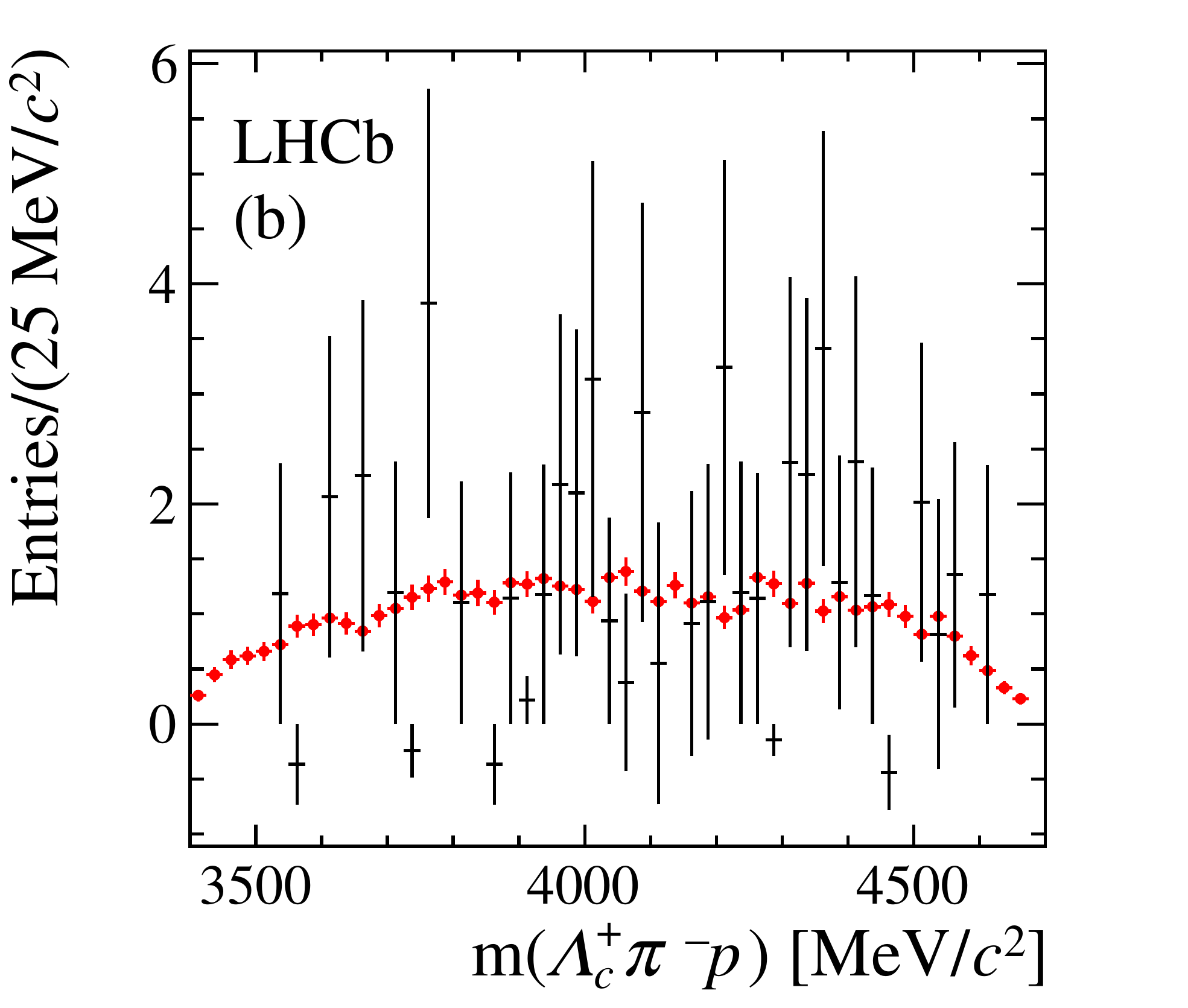}
\includegraphics[width=0.32\textwidth]{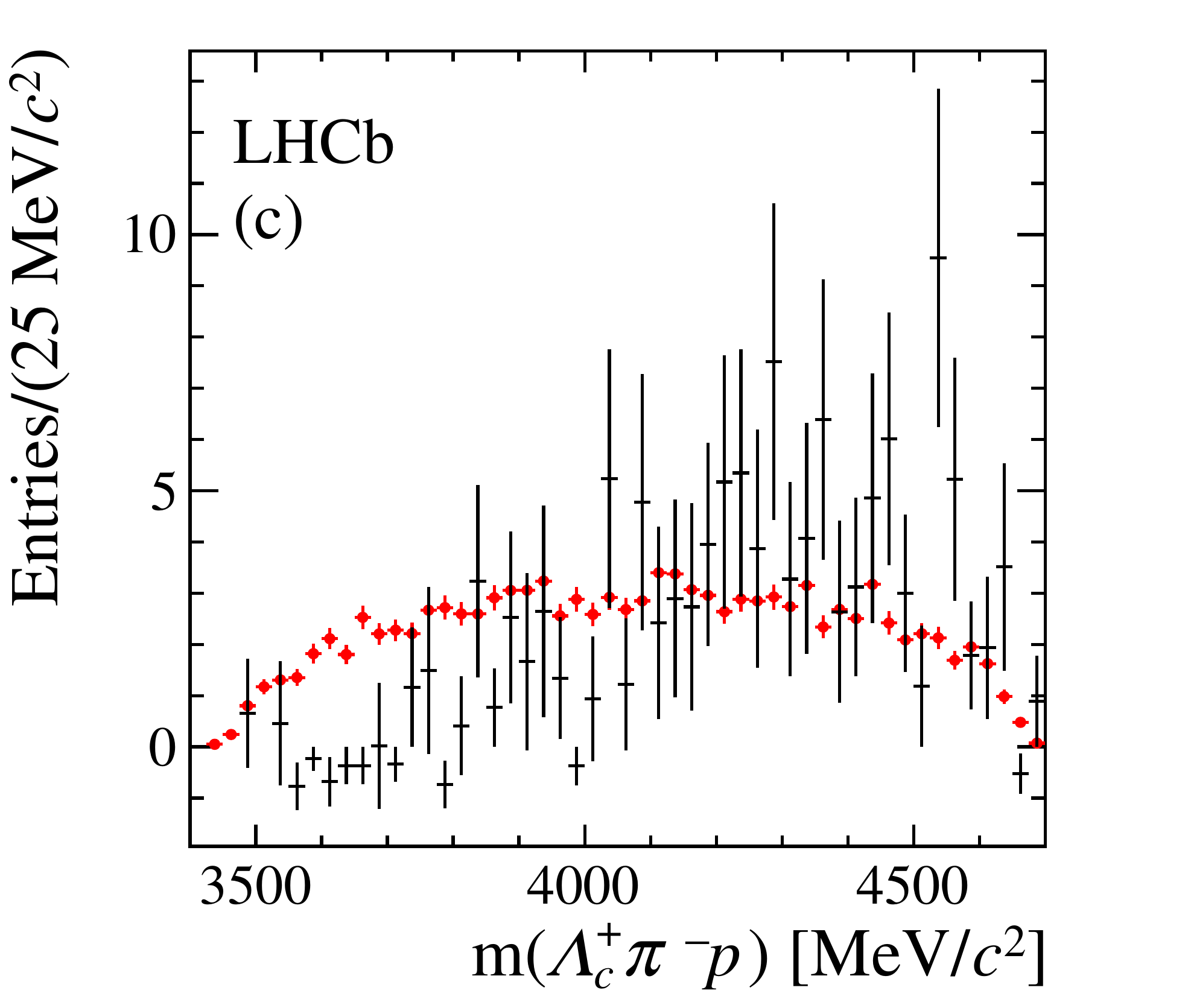}
\caption{Background-subtracted mass spectrum of the $\Lc\pim\proton$ system from the decay
${\Lb\to\Lc\proton\antiproton\pim}$ in (a) the full $\Lc\pim$ mass spectrum,
(b) the  signal region of the $\PSigma_c^0$ resonance, and (c) 
  the signal region of the $\PSigma_c^{*0}$ resonance. In all figures, the
  black points are data and the red points  
  are simulated events where the \Lb baryon decays to the \Lc\proton\antiproton\pim final state 
  (a) based on a uniform-phase-space model, (b) through the $\PSigma_c^0$ resonance and (c) 
  through the $\PSigma_c^{*0}$ resonance.  
  No evident peaking shapes are visible.}
\label{fig:dibaryon}
\end{center}
\end{figure}

The existence of dibaryon resonances, 
${\mathscr{D}_c^+\to\proton\PSigma_c^0}$, is investigated in the 
${\Lc\pim\proton}$ mass spectrum of background-subtracted data.  
The full $m(\Lc\pim)$ spectrum is considered, while the signal regions of 
$\PSigma_c^0$ and $\PSigma_c^{*0}$ resonances are defined  by the ranges 
${ 2450 < m(\Lc\pim) <  2458\mevcc }$ and  ${2488 < m(\Lc\pim) <  2549\mevcc}$, respectively.   
The background is subtracted with the \sPlot technique\cite{Pivk:2004ty}.
No peaking structures are observed in the 
distributions shown in Figure~\ref{fig:dibaryon}. 
The two-dimensional distribution of $m(\Lc\proton\pim)$ versus $m(\Lc\pim)$ 
has been checked and does not exhibit  
any clear structure. 

\section{Conclusion}
The first observation of the decay ${\Lb\to\Lc\proton\antiproton\pim}$ is presented. The
ratio of the branching fractions using the decay ${\Lb \to \Lc \pim}$ 
as the  normalisation  channel  
is measured to be
\begin{equation*}
    \frac{\mathcal{B}(\Lb \to \Lc \proton \antiproton \pim)}{\mathcal{B}(\Lb \to \Lc \pim)} = 0.0540 \pm 0.0023 \pm 0.0032,
\end{equation*}
using data corresponding to an integrated luminosity of 3\invfb collected during 2011 and 2012 with the \lhcb detector.
Contributions from the $\PSigma_c(2455)^0$ and $\PSigma_c^{*}(2520)^0$ resonances are observed,
and the ratios of their branching fractions with respect to the ${\Lb \to \Lc\proton \antiproton  \pim}$ decays  are measured to be
\begin{align*}
  \frac{\mathcal{B}(\Lb \to \PSigma_c^0 \proton \antiproton )\times\mathcal{B}(\PSigma_c^0\to \Lc \pim)}{\mathcal{B}(\Lb \to \Lc  \proton \antiproton \pim)} = 0.089\pm0.015\pm0.006,\\
  \frac{\mathcal{B}(\Lb \to \PSigma_c^{*0} \proton \antiproton )\times\mathcal{B}(\PSigma_c^{*0}\to \Lc \pim)}{\mathcal{B}(\Lb \to \Lc\proton \antiproton\pim)} = 0.119\pm0.020\pm0.014.
\end{align*}
In all of the above results, the first uncertainty is statistical and the second is systematic.

The mass spectra of the $\Lc\proton\pim$ final state are also inspected for possible dibaryon
resonances, but no evidence of peaking structures is observed.

\section*{Acknowledgements}
%
%
\noindent We express our gratitude to our colleagues in the CERN
accelerator departments for the excellent performance of the LHC. We
thank the technical and administrative staff at the LHCb
institutes. We acknowledge support from CERN and from the national
agencies: CAPES, CNPq, FAPERJ and FINEP (Brazil); MOST and NSFC
(China); CNRS/IN2P3 (France); BMBF, DFG and MPG (Germany); INFN
(Italy); NWO (The Netherlands); MNiSW and NCN (Poland); MEN/IFA
(Romania); MinES and FASO (Russia); MinECo (Spain); SNSF and SER
(Switzerland); NASU (Ukraine); STFC (United Kingdom); NSF (USA).  We
acknowledge the computing resources that are provided by CERN, IN2P3
(France), KIT and DESY (Germany), INFN (Italy), SURF (The
Netherlands), PIC (Spain), GridPP (United Kingdom), RRCKI and Yandex
LLC (Russia), CSCS (Switzerland), IFIN-HH (Romania), CBPF (Brazil),
PL-GRID (Poland) and OSC (USA). We are indebted to the communities
behind the multiple open-source software packages on which we depend.
Individual groups or members have received support from AvH Foundation
(Germany), EPLANET, Marie Sk\l{}odowska-Curie Actions and ERC
(European Union), ANR, Labex P2IO and OCEVU, and R\'{e}gion
Auvergne-Rh\^{o}ne-Alpes (France), Key Research Program of Frontier
Sciences of CAS, CAS PIFI, and the Thousand Talents Program (China),
RFBR, RSF and Yandex LLC (Russia), GVA, XuntaGal and GENCAT (Spain),
Herchel Smith Fund, the Royal Society, the English-Speaking Union and
the Leverhulme Trust (United Kingdom).


\appendix

\addcontentsline{toc}{section}{References}
\setboolean{inbibliography}{true}
\bibliographystyle{LHCb}
\bibliography{main,LHCb-PAPER,LHCb-CONF,LHCb-DP,LHCb-TDR}

\newpage

\centerline{\large\bf LHCb collaboration}
\begin{flushleft}
\small
R.~Aaij$^{43}$,
B.~Adeva$^{39}$,
M.~Adinolfi$^{48}$,
Z.~Ajaltouni$^{5}$,
S.~Akar$^{59}$,
P.~Albicocco$^{19}$,
J.~Albrecht$^{10}$,
F.~Alessio$^{40}$,
M.~Alexander$^{53}$,
A.~Alfonso~Albero$^{38}$
% ...

{\footnotesize \it
% ...
$ ^{5}$Clermont Universit{\'e}, Universit{\'e} Blaise Pascal, CNRS/IN2P3, LPC, Clermont-Ferrand, France\\
% ...
$ ^{10}$Fakult{\"a}t Physik, Technische Universit{\"a}t Dortmund, Dortmund, Germany\\
% ...
$ ^{19}$Laboratori Nazionali dell'INFN di Frascati, Frascati, Italy\\
% ...
$ ^{38}$ICCUB, Universitat de Barcelona, Barcelona, Spain\\
$ ^{39}$Instituto Galego de F{\'\i}sica de Altas Enerx{\'\i}as (IGFAE), Universidade de Santiago de Compostela, Santiago de Compostela, Spain\\
$ ^{40}$European Organization for Nuclear Research (CERN), Geneva, Switzerland\\
% ...
$ ^{43}$Nikhef National Institute for Subatomic Physics, Amsterdam, The Netherlands\\
% ...
$ ^{48}$H.H. Wills Physics Laboratory, University of Bristol, Bristol, United Kingdom\\
% ...
$ ^{53}$School of Physics and Astronomy, University of Glasgow, Glasgow, United Kingdom\\
% ...
$ ^{59}$University of Cincinnati, Cincinnati, OH, United States\\
% ...
}
\end{flushleft}

\newpage

\end{document}